\documentclass[twocolumn,secnumarabic,amssymb, nobibnotes, aps, prb,groupedaddress, superscriptaddress]{revtex4-2}
\usepackage{amsmath,graphicx,latexsym,times,color}
\usepackage{setspace}
\usepackage[hidelinks,colorlinks=true, allcolors=blue]{hyperref}
\usepackage{array}
\usepackage{textcomp}
\usepackage{titlesec}
\usepackage{physics}
\usepackage{gensymb}
\usepackage{nccmath}
\usepackage{empheq} 
\usepackage[normalem]{ulem} 
\usepackage{float}

\definecolor{blue-violet}{rgb}{0.54, 0.17, 0.89} 

\let\oldtimes\times  
\renewcommand\times{{\oldtimes}}

\usepackage[cmyk,dvipsnames]{xcolor}
\definecolor{darkorchid}{HTML}{bf3eff}

\begin{document}
    \title{Universal two-stage dynamics and phase control in skyrmion formation}
    \author{Shiwei Zhu}
    \author{Xinyuan Guan}
    \author{Zhen Sun}
    \author{Qiuyao Zhang}

\author{Changsheng Song}
\email[Contact author: ]{cssong@zstu.edu.cn}

\affiliation{Department of Physics,
\href{https://ror.org/03893we55}{Zhejiang Sci-Tech University}, Hangzhou 310018, China}

\affiliation{Zhejiang Key Laboratory of Quantum State Control and Optical Field Manipulation, \href{https://ror.org/03893we55}{Zhejiang Sci-Tech University}, Hangzhou 310018, China}

\date{\today}
\begin{abstract}


We uncover a universal two-stage dynamics during skyrmion formation and establish its connection to equilibrium phases through the introduction of a Dzyaloshinskii–Moriya interaction (DMI) chiral correlation $\chi$. Stage I involves stripe coarsening governed by the exchange-to-DMI ratio $J'$, while stage II entails stripe contraction driven by the synergy between $J'$ and the anisotropy-to-DMI ratio $K'$. The magnetic field-to-DMI ratio $B'$ influences both stages. By combining symbolic regression with neural networks, we model the competition and cooperation among these parameters and derive a skyrmion formation criterion, $0.58\,J'K' + \mu B'J' > 1$. Our model disentangles their distinct roles: $J'$ sets the stripe width, $K'$ primarily controls the skyrmion size, and $B'$ strongly affects the topological charge. This approach provides a general framework for predicting and controlling magnetic phases in chiral magnets.



\end{abstract}

\maketitle

Magnetic skyrmions are topologically protected nanoscale spin textures~\cite{skyrme1962unified} that exhibit particle-like properties and hold great promise for low-power spintronic applications~\cite{ fert2013skyrmions, fert2017magnetic, back20202020, gobel2021beyond, morshed2022positional, he2023all}. These structures emerge from the competition between chiral and symmetric spin interactions at the microscopic level. The Dzyaloshinskii–Moriya interaction (DMI) favors a fixed chirality by promoting a perpendicular sense of spin rotation between neighboring spins, thereby stabilizing noncollinear spin textures~\cite{DZYALOSHINSKY1958241, PhysRev.120.91}. In contrast, the ferromagnetic Heisenberg exchange $J$ favors parallel spin alignment, counteracting the DMI-induced canting~\cite{mermin1966absence}. Magnetic anisotropy $K$ constrains the spin orientation along a specific easy axis~\cite{bander1988ferromagnetism, wilson2014chiral}, while an external magnetic field $B$ couples to the spins via the Zeeman interaction, inducing global polarization along the field direction~\cite{ bogdanov1994thermodynamically, zhang2018manipulation}. The delicate balance among these competing interactions governs the formation of various topological magnetic phases, including labyrinthine stripes, isolated skyrmions, and skyrmion lattices, which have been experimentally observed in a wide range of systems, from bulk crystals and ultrathin films to two-dimensional van der Waals materials~\cite{Muhlbauer2009skyrmion, yu2010real, yu2011near, yu2012magnetic, woo2016observation, cortes2019nanoscale, zhang2023direct, ding2019observation, zhang2022room}.

While the equilibrium properties of chiral magnetic phases have been well characterized~\cite{roessler2006spontaneous, leonov2016properties, wijethunga2025phase, wang2018theory, wu2021size, hu2022theory, wu2022nematic}, the dynamical processes governing skyrmion formation remain poorly understood. Fundamental questions persist regarding how these competing interactions collectively drive the real-time evolution of spin systems toward topological states, and whether universal principles underlie chiral magnetic state formation across different parameter regimes. Understanding these nonequilibrium dynamics is essential for revealing the energy landscape that controls topological phase transitions and for achieving deterministic control of metastable states in practical device applications.

Addressing these challenges requires confronting the high dimensionality of spin dynamics and the nonlinear coupling among multiple parameters. Machine learning (ML) approaches offer particular promise for this task, having demonstrated powerful capabilities in extracting features from complex data~\cite{feng2024classification, hu2025machine}, accelerating multiscale simulations~\cite{tang2023neural,li2024neural}, and identifying phase transitions~\cite{carrasquilla2017machine,wang2021learning, xiong2025capturing}. The ability of these methods to detect hidden correlations in large datasets makes them especially well suited for deciphering the nonequilibrium evolution of chiral magnetic systems.

In this Letter, we combine large-scale spin dynamics simulations with ML to reveal how magnetic parameters collectively govern not only the equilibrium properties but also the nonequilibrium dynamics from initially disordered spins to chiral magnetic states. We introduce a normalized chiral correlation $\chi$ as an order parameter, whose rise and decay uncover a universal two-stage mechanism: initial stripe coarsening followed by subsequent contraction. By integrating symbolic regression with neural networks, we establish a generic relation valid across both labyrinthine stripe and skyrmion states, revealing the competition and cooperation among magnetic parameters. Phase diagrams identified by neural networks with 98\% accuracy confirm the parameter dependence of $\chi$ in different magnetic states and yield the finite-field boundary for skyrmion formation, $0.58\,J'K' + \mu B'J' = 1$. Further predictions, together with local energy analyses, clarify how magnetic parameters control phase stability. Our results not only provide a microscopic understanding of skyrmion formation but also demonstrate how decoding nonequilibrium dynamics enables predictive control of topological phases, offering a new paradigm for designing functional magnetic materials.

To investigate spin textures under various conditions, we use the following spin Hamiltonian to map the total energy:

\begin{equation}\label{hamiltonian}
    \begin{split}
        H & =-\sum_{\langle i,j \rangle}J_{ij}(\mathbf{S}_i \cdot \mathbf{S}_j)-\sum_{\langle i,j \rangle}\mathbf{D}_{ij} \cdot(\mathbf{S}_i \times \mathbf{S}_j) \\&\hphantom{=\ } -K\sum_i (S_i^z)^2 -\mu B\sum_i (S_i^z),
    \end{split}
\end{equation}
where $J_{ij} > 0$ denotes the ferromagnetic Heisenberg exchange between neighboring spins $\mathbf{S}_i$ and $\mathbf{S}_j$.
$\mathbf{D}_{ij}$ is the DMI vector with magnitude $D = |\mathbf{D}_{ij}|$, oriented in the lattice plane and perpendicular to the bond direction.
$K$ represents the easy-axis single-ion anisotropy, and $B$ is an external magnetic field applied along the $z$ axis.
$\mu$ is the magnetic moment of each atom.

Atomistic spin dynamics simulations are performed with \texttt{Spirit} package~\cite{PhysRevB.99.224414}, based on the Landau–Lifshitz–Gilbert equation~\cite{landau1992theory,1353448}.
A $200 \times 200$ lattice with periodic boundary conditions is employed to approximate an infinite system, and each simulation is evolved for 500,000 integration steps to reach a dynamically stabilized configuration.
The simulations are initialized from a disordered spin configuration to trigger the ordering process, and the temperature is set to zero to suppress stochastic fluctuations, effectively corresponding to a zero-temperature quench.
A relatively large damping factor is used to suppress spin precession and highlight the collective ordering dynamics governed by the effective field (see Sec.~S1 of the supplemental material (SM)~\cite{supplmat} for validity tests regarding the choice of damping factor, temperature, and initial configurations).

\begin{figure}[t]
	\centering
	\includegraphics[width=0.9\linewidth]{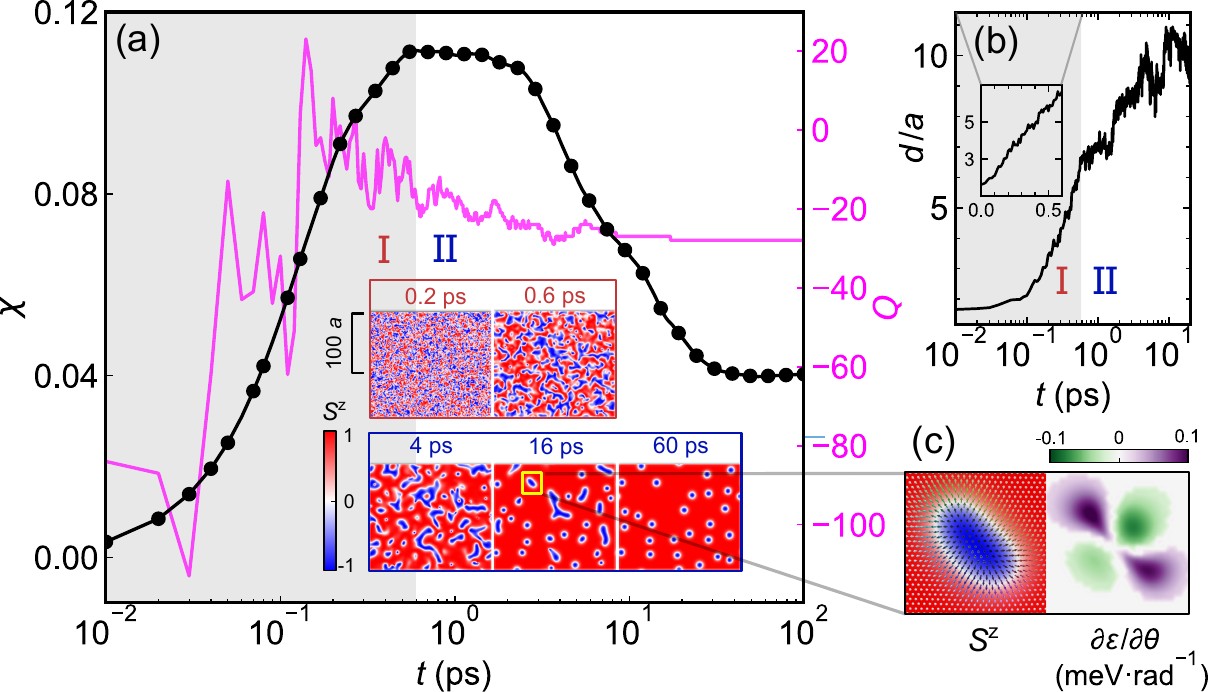}
	\caption{\label{fig_1}
Universal two-stage dynamics of skyrmion formation.
    (a) Evolution of the normalized chiral correlation $\chi$ (black, left axis) and topological charge $Q$ (pink, right axis) on a logarithmic time scale. Stage~I (gray shading) shows coarsening of stripe-like domains, characterized by increasing $\chi$ and oscillatory $Q$, while stage~II exhibits stripe contraction into skyrmions, marked by decreasing $\chi$ and saturation of $Q$. Spin snapshots illustrate the morphological evolution during the two stages.
(b) Evolution of the average stripe width $d$ (all length values are normalized by the lattice constant $a$) on a logarithmic time scale, highlighting near-linear growth in stage~I (inset) and oscillatory deformation in stage~II.
(c) Zoomed-in view of a stripe at 16~ps (stage~II) undergoing rotational contraction into a skyrmion (left). The corresponding map of $\partial \epsilon/\partial\theta$ (right) shows the derivative of the local energy $\epsilon_i$ with respect to the spin polar angle $\theta$, where positive values (purple) indicate the driving torque that aligns spins along the $z$ axis.
    }
\end{figure}

To quantitatively characterize the evolution of spin configurations during the ordering process, we define a normalized chiral correlation as

\begin{equation}\label{chiral}
    \chi=\frac{H_{\text{DMI}}}{{-D} \, N_{\text{pairs}}}=\frac{1}{N_{\text{pairs}}}\sum_{\left\langle i,j\right\rangle}\hat{\mathbf{D}}_{ij}\cdot\left({\mathbf{S}}_{i}\times{\mathbf{S}}_{j}\right),
\end{equation}
where $\hat{\mathbf{D}}_{ij}$ is the unit Dzyaloshinskii–Moriya vector between spins $i$ and $j$, and $N{\mathrm{pairs}}$ is the number of nearest-neighbor atom pairs.
The parameter $\chi$ quantifies the global chiral ordering and is directly linked to the DMI energy, thus capturing both the spin configuration and the underlying energy competition.

The real-time evolution of $\chi$ reveals a universal two-stage dynamical process in skyrmion formation, as illustrated in Fig.~\ref{fig_1}(a). 
In stage~I ($t = 0$–$0.6$~ps), $\chi$ increases monotonically, signaling the coarsening of stripe-like domains (stripes) from the initial disordered state. 
During this stage, the competition between exchange interaction ($E_{\mathrm{ex}} \propto J q^{2}$) and DMI ($E_{\mathrm{DMI}} \propto -D q$) selects a characteristic wave vector $q \propto D/J$~\cite{rohart2013skyrmion}, leading to stripe formation, and a corresponding linear increase in the stripe width $d$ [Fig.~\ref{fig_1}(b), inset].
The topological charge $Q$, reflecting the number of skyrmions, exhibits pronounced oscillations, indicative of transient topological fluctuations prior to skyrmion stabilization.

Stage~II ($t > 0.6$~ps) is characterized by a decrease in $\chi$, as stripes contract into individual skyrmions, while $Q$ gradually saturates. 
This morphological transition arises from the competition among collinear energy terms (exchange, anisotropy, and Zeeman energy) and the chiral DMI energy (see Sec.~S2 of the SM~\cite{supplmat}). 
The contraction mechanism is further clarified by the spatial distribution of $\partial \epsilon / \partial \theta$, the derivative of the local energy with respect to the spin polar angle $\theta$ [Fig.~\ref{fig_1}(c)]. 
Regions with positive $\partial \epsilon / \partial \theta$ indicate a local tendency of spins to align along the $z$~axis, thereby driving contraction along the stripe's long axis.
Simultaneously, a slight expansion along the short axis reflects a rotational deformation during the stripe-to-skyrmion transformation, leading to the slow oscillatory behavior of the stripe width in Fig.~\ref{fig_1}(b).

\begin{figure*}[t]
	\centering
	\includegraphics[width=0.95\linewidth]{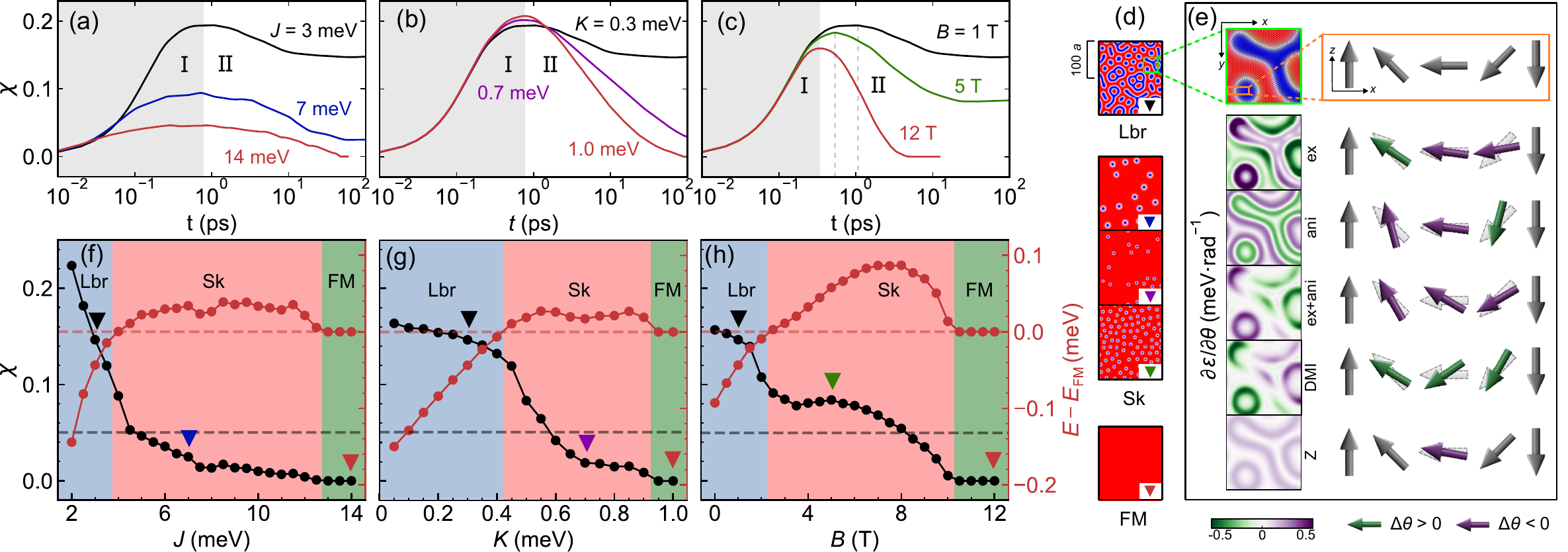}
	\caption{\label{fig_2}
Two-stage dynamics and converged states under parameter variation.
Time evolution of the normalized chiral correlation $\chi$ under variation of (a) exchange interaction $J$, (b) anisotropy $K$, and (c) magnetic field $B$.
The reference parameters are $J=3$, $D=1$, $K=0.3~\mathrm{meV}$, and $B=1~\mathrm{T}$ (black curve in each panel). Grey shading indicates stage I (coarsening) and stage II (contraction). 
(d) Representative snapshots of converged states, marked by symbols: labyrinth domain (Lbr, dark triangle), skyrmion (Sk, blue, purple, and green triangles), and ferromagnetic (FM, red triangle) states.
(e) Influence of individual energy terms on a labyrinth‐domain segment in (d), evaluated via $\partial \epsilon/\partial\theta$ for the exchange (ex), anisotropy (ani), DMI, and Zeeman (Z) energies. The grey arrow and the dashed arrow denote the initial orientations of the spins, while the green and purple arrows denote the rotational tendencies induced by each energy term, driving the spins away from or towards the $z$ axis, respectively.
(f–h) Converged values of $\chi$ and the energy difference relative to the FM state, as functions of $J$, $K$, and $B$.
The red solid circle-lines denote $E - E_\mathrm{FM}$, and the intersections with the red dashed lines define the phase boundaries (Lbr–Sk and Sk–FM). The grey dashed line indicates the significant chiral order with $\chi$ = 0.05.
    }
\end{figure*}

To elucidate the distinct roles of magnetic parameters in skyrmion formation, we systematically vary the exchange coupling $J$, anisotropy $K$, and magnetic field $B$ to track the resulting evolution of the chiral correlation $\chi$. 
As shown in Fig.~\ref{fig_2}(a), increasing $J$ markedly suppresses the peak value of $\chi$ at the end of stage~I and maintains this reduction throughout stage~II, confirming that $J$ primarily governs the initial coarsening dynamics. 
In contrast, varying $K$ has little influence during stage~I but significantly accelerates the decay of $\chi$ in stage~II [Fig.~\ref{fig_2}(b)]. 
Increasing $B$ induces a moderate reduction of $\chi$ at the end of stage~I [gray dashed lines in Fig.~\ref{fig_2}(c)] and enhances its suppression during stage~II, indicating that $B$ affects both stages of evolution.

Through these processes, the system evolves from an initially disordered state into different magnetic configurations depending on the parameters. 
When the collinear terms are weak, stripes in stage~II become more ordered without contraction, forming a labyrinthine domain state (Lbr). 
With intermediate parameters, the stripes gradually contract into skyrmion (Sk) states with different sizes and densities. 
Further increase of the collinear terms causes complete contraction and skyrmion annihilation, yielding a ferromagnetic (FM) state. 
Representative spin configurations are shown in Fig.~\ref{fig_2}(d).

We further analyze the local energy landscape of a labyrinth domain segment by evaluating $\partial \epsilon/\partial\theta$ for each energy term [Figure~\ref{fig_2}(e)]. 
The exchange (ex) term drives spins at the outer wall edge to tilt downward and inner spins upward, with strength modulated by wall curvature, causing convex segments to contract and concave ones to expand. 
The anisotropy (ani) term induces opposing spin rotations, narrowing the domain wall but leaving its position nearly unchanged due to comparable inner and outer strengths.
When $J$ and $K$ act cooperatively (ex+ani), they trigger inward rotation only at stripe terminals, enabling skyrmion shrinkage while keeping straight stripes intact, which explains why $K$ becomes effective mainly in stage~II.  
In contrast, the DMI acts oppositely to the cooperative effect of $J$ and $K$, whereas the Zeeman (Z) term uniformly aligns spins along the $z$~axis, compressing both stripe and skyrmion textures. Together, these two terms also promote stripe fragmentation.

The converged chiral correlation $\chi$ and the energy difference relative to the FM state, plotted as functions of $J$, $K$, and $B$ in Figs.~\ref{fig_2}(f–h), classify the Lbr, Sk, and FM phases.
The exchange coupling $J$ exerts the strongest influence on both $\chi$ and the relative energy within the Lbr phase.
The anisotropy $K$ has little effect in the Lbr phase but markedly modifies $\chi$ in the Sk phase.
The magnetic field $B$ maintains a plateau in $\chi$ just beyond the Lbr–Sk transition, expanding the region with $\chi > 0.05$, which corresponds to slightly higher energies than typical Sk states shown in Figs.~\ref{fig_2}(f–g).

\begin{figure}[t]
	\centering
	\includegraphics[width=0.765\linewidth]{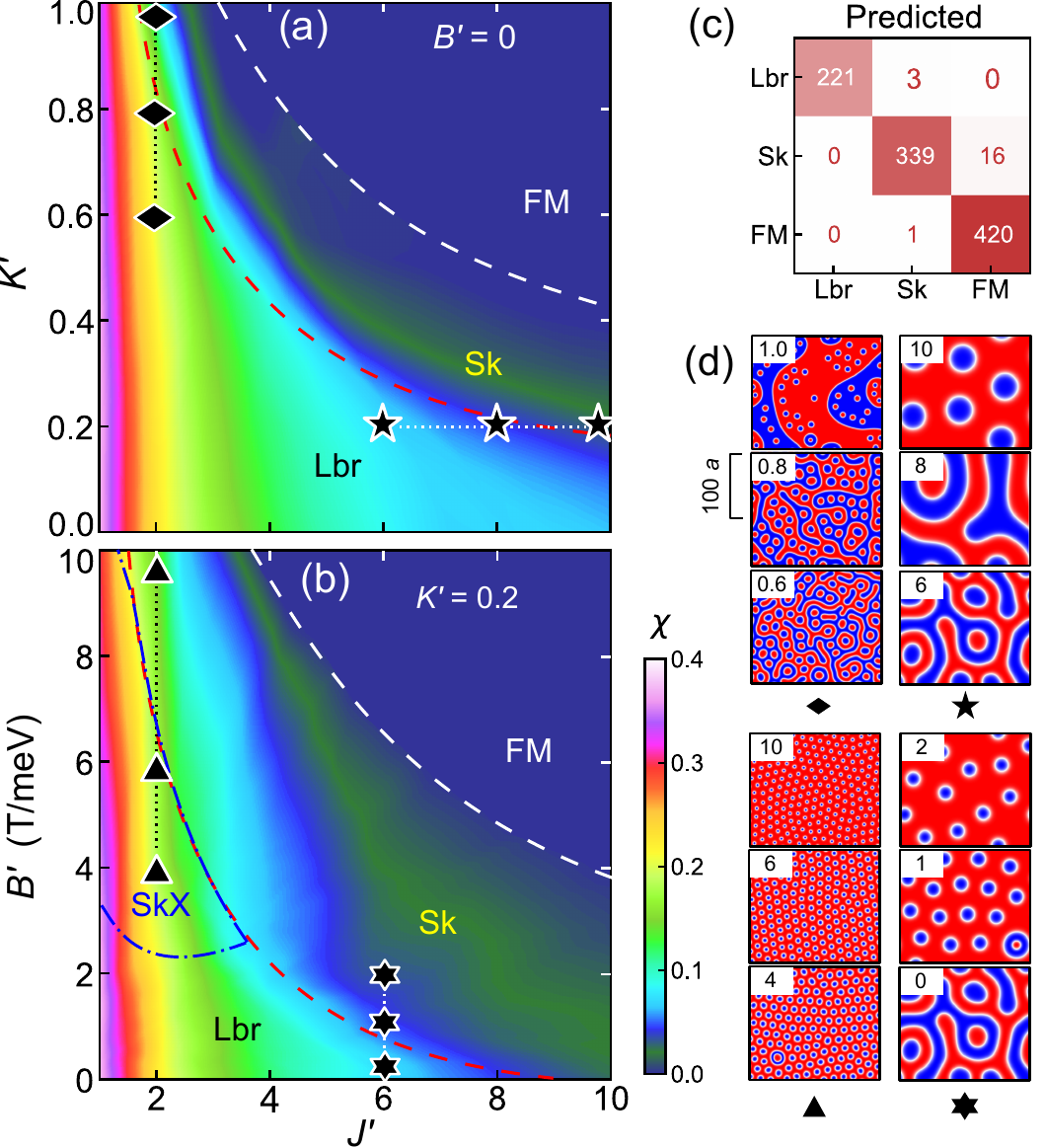}
	\caption{\label{fig_3}
Machine-learning prediction of magnetic phase diagrams and skyrmion morphology control.
(a) Zero-field phase diagram in the $J'–K'$ parameter space. 
Phase boundaries are marked by red dashed lines (Lbr–Sk) and white dashed lines (Sk–FM). 
The color scale indicates the chiral correlation $\chi$. 
(b) Phase diagram in the $J'–B'$ plane at fixed $K' = 0.2$. 
The dot–dash line marks the skyrmion lattice (SkX) region stabilized at low $J'$ and intermediate $B'$.
(c) Confusion matrix of the phase-classification model, achieving 98\% accuracy across Lbr, Sk, and FM phases.
(d) Spin configurations corresponding to labeled regions in (a) and (b): sparse stripes and small skyrmions with large $K'$ (0.6, 0.8, and 1.0) at low $J'$ (diamond); wide stripes and large skyrmions with low $K'$ (6, 8, and 10) at high $J'$ (star); SkX remains stable with moderate $B'$ (4, 6, and 10) at low $J'$ (triangle), but becomes suppressed with low $B'$ (0, 1, 2) at higher $J'$ (hexagram).
    }
\end{figure}
To establish a quantitative link between microscopic parameters and magnetic phases, we apply symbolic regression using the Python library \texttt{PySR}~\cite{cranmer2023interpretable}, which identifies interpretable functional forms from candidate terms and operators through evolutionary algorithms.
For simplicity, Eq.~\ref{hamiltonian} is normalized by the DMI constant $D$, rendering the energy terms dimensionless. We then define $J' = J/D$, $K' = K/D$, and $B' = B/D$ to characterize the competition among these parameters.

By considering the decaying effect of those terms on $\chi$ as an exponential factor, the two-stage dynamics can be compactly written as

\begin{equation}\label{chi}
\chi \sim\exp[-f_0(\kappa_1)] \cdot\exp[-f_1(\kappa_1) f_2(\kappa_2) f_3(\kappa_3)],
\end{equation}
with $\kappa_1 = \sqrt{J'}$, $\kappa_2 = (J'K')^3 K'$, $\kappa_3 = B'$, and each $f_i$ being a positive linear function (see Sec.~S3 of the SM for details).
This expression highlights the dominant role of $\kappa_1$ (the square-root dependence on $J'$) when the DMI is strong, whereas the higher-order term $\kappa_2$, representing the cooperative effects of $J'$ and $K'$, becomes significant when the DMI is weaker.
The linear dependence of $\kappa_3$ on $B'$ enables the magnetic field to uniformly modulate $\chi$.
Symbolic regression achieves $R^2 > 98\%$, confirming the reliability of the obtained relation. Importantly, this factorization highlights the two-stage dynamics: the first term, $\exp[-f_0(\kappa_1)]$, characterizes $\chi$ at the end of stage I, while the second term, $\exp[-f_1(\kappa_1) f_2(\kappa_2) f_3(\kappa_3)]$, describes its evolution in stage II, showing how different magnetic parameters govern each stage.



To verify the trend revealed by Eq.~(\ref{chi}), we further validate the phase behavior using a neural network, which predicts the phase diagrams with 98\% accuracy, as shown by the confusion matrix in Fig.~\ref{fig_3}(c). The zero-field phase diagram [Fig.~\ref{fig_3}(a)] reveals distinct regions of Lbr, Sk and FM phases, with color indicating the chiral correlation $\chi$. Phase boundaries are determined symbolically, with approximate form: $0.58\,J'K'>1$ for Lbr-Sk boundary. This scaling form agrees with earlier analyses~\cite{wang2018theory} which suggested $JK/D^2 (J^\prime K^\prime)$ as a key stability parameter.
In the stripe phase, as indicated by nearly vertical spectral contours, showing that stripe formation is governed mainly by $J'$. In contrast, the reciprocal-like contour in the skyrmion phase reveals a $J'K'$ dependence, consistent with the trend described by Eq.~(\ref{chi}). The variety of $\chi$ across the boundary reflects distinct boundary states.
Representative spin configurations in Fig.~\ref{fig_3}(d) illustrate that increasing $J'$ (star in Fig.~\ref{fig_3}(a)) drives a transition from fine to wide stripes and eventually to large skyrmions, while increasing $K'$ (diamond) maintains the stripe but results in smaller skyrmions after the transition.

Figure~\ref{fig_3}(b) presents the phase diagram in the $J'$–$B'$ plane at fixed $K'=0.2$.
At low $J'$, $\chi$ remains high ($0.05$–$0.2$) due to the $J'$ dependence for Lbr state. Increasing $B'$ (triangle) generates more skyrmions and stabilizes an extended SkX region, consistent with the skyrmion-number dependence~\cite{wang2023topological}. Even after entering the Sk state at high fields, the system maintains a large skyrmion number over a wide field range (e.g., at $B=10$ T).
Conversely, at high $J'$ (hexagram), the skyrmion number is much lower, and the SkX order is suppressed.
Extending the symbolic regression to include $B'$ yields an approximate field-dependent relation for the Lbr–Sk boundary,
\begin{equation}\label{boundary}
0.58\,J'K' + \mu B'J' = 1,
\end{equation}
which is also suitable for the zero-field case (see Sec.~S4 in the SM for the full expression~\cite{supplmat}).
Combining the fitted phase boundaries with the $\chi$ landscape provides a comprehensive understanding of the effects of magnetic parameters on skyrmion formation and stability.

\begin{figure}[t]
	\centering
	\includegraphics[width=0.95\linewidth]{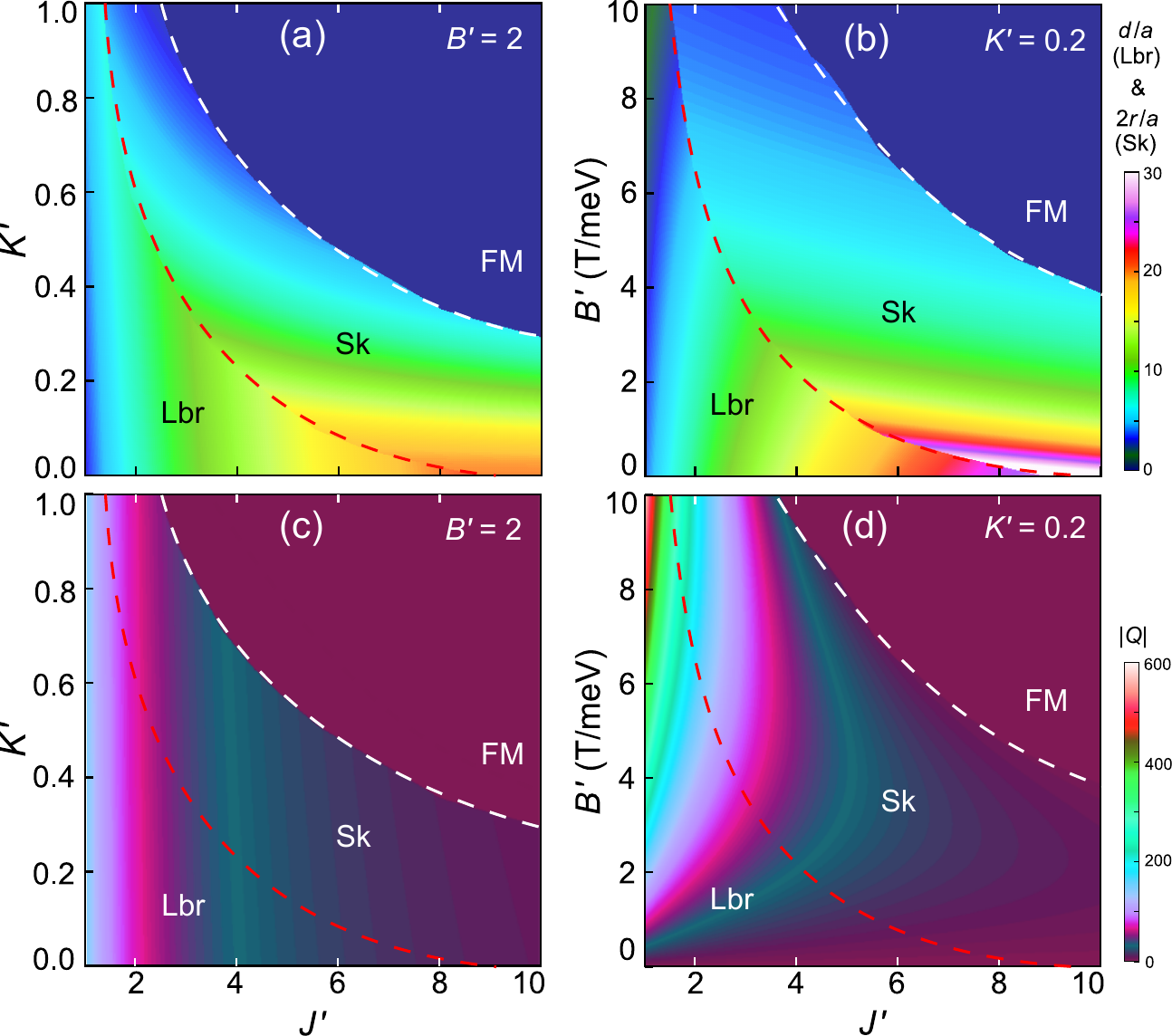}
	\caption{\label{fig_4}
Symbolic regression predicted magnetic properties across parameter space. Color scale indicates magnitude.
Stripe width $d$ (in Lbr phase) and skyrmion radius $r$ (in Sk phase) plotted in the (a) $J'$–$K'$ space at $B'=2$~T/meV and in the (b) $J'$–$B'$ space with $K'=0.2$, respectively.
(c, d) Corresponding absolute topological charge $|Q|$ under the same respective conditions.
    }
\end{figure}

Figure~\ref{fig_4} summarizes the symbolic-regression predictions for stripe and skyrmion properties (see Sec.~S3 and S5 of the SM~\cite{supplmat} for details). 
In panel~(a), the stripe width $d$ is primarily controlled by $J'$, whereas the skyrmion radius $r$ is more sensitive to $K'$. 
As shown in panel~(b), $B'$ exhibits a similar effect to $K'$, but also exerts a slight influence on the stripe width, consistent with our analysis of Fig.~\ref{fig_2}(e). 
One can notice that only the high-$K'$ region crosses the Sk–FM boundary with a small skyrmion radius. 
The minimum skyrmion radius is found to scale with the skyrmion wall width $w$, which is strongly influenced by $K'$ (see Sec.~S6 of the SM~\cite{supplmat}). 
Therefore, achieving extremely small skyrmions for applications may require a larger $K'$.

Another important property for applications is the topological charge $Q$, which determines the skyrmion number and density. 
As shown in Fig.~\ref{fig_4}(c), increasing $J'$ rapidly reduces the absolute topological charge $|Q|$. 
Panel~(d) further demonstrates a dual effect of the magnetic field on $|Q|$: it first increases sharply with increasing field, then decreases in the later stage of the skyrmion phase, and finally approaches zero near the FM transition.
It is worth noting that the topological charge and the stripe width exhibit similar trends, reflecting their common origin: finer stripes in stage~I result in more small magnetic domains.
Moreover, the magnetic field also promotes stripe fragmentation, further enhancing $|Q|$. 
This makes the magnetic field a crucial factor for SkX formation and provides an approach to predict its phase region (see Sec.~S6 of the SM~\cite{supplmat}).

In summary, our study reveals a universal two-stage dynamic of skyrmion formation, clarifying how competing magnetic interactions collectively control topological phase formation. 
During stage~I, initial stripe coarsening determines the stripe width and the topological charge under the influence of $J'$ and $B'$. 
In stage~II, stripe-like domains rotate and contract, governed by the cooperative effect of $J'$ and $K'$ and the independent effect of $B'$. 
By combining symbolic regression with neural-network analysis, we establish quantitative scaling relations for chiral-state formation and field-dependent phase stability boundaries. 
This framework clearly disentangles the distinct roles of key parameters: $J'$ primarily sets the stripe width and thus affects the topological charge, $K'$ mainly governs the skyrmion size, and $B'$ first increases the skyrmion number to stabilize a SkX state, then decreases it. 
Overall, these results clarify the microscopic mechanism of skyrmion formation and establish a predictive framework for phase control. 
Furthermore, this work demonstrates the effectiveness of learning nonequilibrium dynamics to interpret equilibrium relations.

 {\textbf{Acknowledgments.}} This work was supported by National Natural Science Foundation of China (Grant No.11804301), the Natural Science Foundation of Zhejiang Province (Grant No. LMS25A040001)

 {\textbf{Data Availability.}} The datasets generated and analyzed in this study are not publicly available due to their large size and the associated cost of storage and hosting. The data are available from the corresponding author upon reasonable request.
	
\bibliography{References}

\begin{thebibliography}{44}%
\makeatletter
\providecommand \@ifxundefined [1]{%
 \@ifx{#1\undefined}
}%
\providecommand \@ifnum [1]{%
 \ifnum #1\expandafter \@firstoftwo
 \else \expandafter \@secondoftwo
 \fi
}%
\providecommand \@ifx [1]{%
 \ifx #1\expandafter \@firstoftwo
 \else \expandafter \@secondoftwo
 \fi
}%
\providecommand \natexlab [1]{#1}%
\providecommand \enquote  [1]{``#1''}%
\providecommand \bibnamefont  [1]{#1}%
\providecommand \bibfnamefont [1]{#1}%
\providecommand \citenamefont [1]{#1}%
\providecommand \href@noop [0]{\@secondoftwo}%
\providecommand \href [0]{\begingroup \@sanitize@url \@href}%
\providecommand \@href[1]{\@@startlink{#1}\@@href}%
\providecommand \@@href[1]{\endgroup#1\@@endlink}%
\providecommand \@sanitize@url [0]{\catcode `\\12\catcode `\$12\catcode
  `\&12\catcode `\#12\catcode `\^12\catcode `\_12\catcode `\%12\relax}%
\providecommand \@@startlink[1]{}%
\providecommand \@@endlink[0]{}%
\providecommand \url  [0]{\begingroup\@sanitize@url \@url }%
\providecommand \@url [1]{\endgroup\@href {#1}{\urlprefix }}%
\providecommand \urlprefix  [0]{URL }%
\providecommand \Eprint [0]{\href }%
\providecommand \doibase [0]{https://doi.org/}%
\providecommand \selectlanguage [0]{\@gobble}%
\providecommand \bibinfo  [0]{\@secondoftwo}%
\providecommand \bibfield  [0]{\@secondoftwo}%
\providecommand \translation [1]{[#1]}%
\providecommand \BibitemOpen [0]{}%
\providecommand \bibitemStop [0]{}%
\providecommand \bibitemNoStop [0]{.\EOS\space}%
\providecommand \EOS [0]{\spacefactor3000\relax}%
\providecommand \BibitemShut  [1]{\csname bibitem#1\endcsname}%
\let\auto@bib@innerbib\@empty
\bibitem [{\citenamefont {Skyrme}(1962)}]{skyrme1962unified}%
  \BibitemOpen
  \bibfield  {author} {\bibinfo {author} {\bibfnamefont {T.~H.~R.}\
  \bibnamefont {Skyrme}},\ }\bibfield  {title} {\bibinfo {title} {A unified
  field theory of mesons and baryons},\ }\href
  {https://doi.org/https://doi.org/10.1016/0029-5582(62)90775-7} {\bibfield
  {journal} {\bibinfo  {journal} {Nucl. Phys.}\ }\textbf {\bibinfo {volume}
  {31}},\ \bibinfo {pages} {556} (\bibinfo {year} {1962})}\BibitemShut
  {NoStop}%
\bibitem [{\citenamefont {Fert}\ \emph {et~al.}(2013)\citenamefont {Fert},
  \citenamefont {Cros},\ and\ \citenamefont {Sampaio}}]{fert2013skyrmions}%
  \BibitemOpen
  \bibfield  {author} {\bibinfo {author} {\bibfnamefont {A.}~\bibnamefont
  {Fert}}, \bibinfo {author} {\bibfnamefont {V.}~\bibnamefont {Cros}},\ and\
  \bibinfo {author} {\bibfnamefont {J.}~\bibnamefont {Sampaio}},\ }\bibfield
  {title} {\bibinfo {title} {Skyrmions on the track},\ }\href
  {https://doi.org/10.1038/nnano.2013.29} {\bibfield  {journal} {\bibinfo
  {journal} {Nat. Nanotechnol.}\ }\textbf {\bibinfo {volume} {8}},\ \bibinfo
  {pages} {152} (\bibinfo {year} {2013})}\BibitemShut {NoStop}%
\bibitem [{\citenamefont {Fert}\ \emph {et~al.}(2017)\citenamefont {Fert},
  \citenamefont {Reyren},\ and\ \citenamefont {Cros}}]{fert2017magnetic}%
  \BibitemOpen
  \bibfield  {author} {\bibinfo {author} {\bibfnamefont {A.}~\bibnamefont
  {Fert}}, \bibinfo {author} {\bibfnamefont {N.}~\bibnamefont {Reyren}},\ and\
  \bibinfo {author} {\bibfnamefont {V.}~\bibnamefont {Cros}},\ }\bibfield
  {title} {\bibinfo {title} {Magnetic skyrmions: advances in physics and
  potential applications},\ }\href {https://doi.org/10.1038/natrevmats.2017.31}
  {\bibfield  {journal} {\bibinfo  {journal} {Nat. Rev. Mater.}\ }\textbf
  {\bibinfo {volume} {2}},\ \bibinfo {pages} {1} (\bibinfo {year}
  {2017})}\BibitemShut {NoStop}%
\bibitem [{\citenamefont {Back}\ \emph {et~al.}(2020)\citenamefont {Back},
  \citenamefont {Cros}, \citenamefont {Ebert}, \citenamefont {Everschor-Sitte},
  \citenamefont {Fert}, \citenamefont {Garst}, \citenamefont {Ma},
  \citenamefont {Mankovsky}, \citenamefont {Monchesky}, \citenamefont
  {Mostovoy} \emph {et~al.}}]{back20202020}%
  \BibitemOpen
  \bibfield  {author} {\bibinfo {author} {\bibfnamefont {C.}~\bibnamefont
  {Back}}, \bibinfo {author} {\bibfnamefont {V.}~\bibnamefont {Cros}}, \bibinfo
  {author} {\bibfnamefont {H.}~\bibnamefont {Ebert}}, \bibinfo {author}
  {\bibfnamefont {K.}~\bibnamefont {Everschor-Sitte}}, \bibinfo {author}
  {\bibfnamefont {A.}~\bibnamefont {Fert}}, \bibinfo {author} {\bibfnamefont
  {M.}~\bibnamefont {Garst}}, \bibinfo {author} {\bibfnamefont
  {T.}~\bibnamefont {Ma}}, \bibinfo {author} {\bibfnamefont {S.}~\bibnamefont
  {Mankovsky}}, \bibinfo {author} {\bibfnamefont {T.}~\bibnamefont
  {Monchesky}}, \bibinfo {author} {\bibfnamefont {M.}~\bibnamefont {Mostovoy}},
  \emph {et~al.},\ }\bibfield  {title} {\bibinfo {title} {The 2020 skyrmionics
  roadmap},\ }\href {https://doi.org/10.1088/1361-6463/ab8418} {\bibfield
  {journal} {\bibinfo  {journal} {J. Phys. D:Appl. Phys.}\ }\textbf {\bibinfo
  {volume} {53}},\ \bibinfo {pages} {363001} (\bibinfo {year}
  {2020})}\BibitemShut {NoStop}%
\bibitem [{\citenamefont {G{\"o}bel}\ \emph {et~al.}(2021)\citenamefont
  {G{\"o}bel}, \citenamefont {Mertig},\ and\ \citenamefont
  {Tretiakov}}]{gobel2021beyond}%
  \BibitemOpen
  \bibfield  {author} {\bibinfo {author} {\bibfnamefont {B.}~\bibnamefont
  {G{\"o}bel}}, \bibinfo {author} {\bibfnamefont {I.}~\bibnamefont {Mertig}},\
  and\ \bibinfo {author} {\bibfnamefont {O.~A.}\ \bibnamefont {Tretiakov}},\
  }\bibfield  {title} {\bibinfo {title} {Beyond skyrmions: Review and
  perspectives of alternative magnetic quasiparticles},\ }\href
  {https://doi.org/10.1016/j.physrep.2020.10.001} {\bibfield  {journal}
  {\bibinfo  {journal} {Phys. Rep.}\ }\textbf {\bibinfo {volume} {895}},\
  \bibinfo {pages} {1} (\bibinfo {year} {2021})}\BibitemShut {NoStop}%
\bibitem [{\citenamefont {Morshed}\ \emph {et~al.}(2022)\citenamefont
  {Morshed}, \citenamefont {Vakili},\ and\ \citenamefont
  {Ghosh}}]{morshed2022positional}%
  \BibitemOpen
  \bibfield  {author} {\bibinfo {author} {\bibfnamefont {M.~G.}\ \bibnamefont
  {Morshed}}, \bibinfo {author} {\bibfnamefont {H.}~\bibnamefont {Vakili}},\
  and\ \bibinfo {author} {\bibfnamefont {A.~W.}\ \bibnamefont {Ghosh}},\
  }\bibfield  {title} {\bibinfo {title} {Positional stability of skyrmions in a
  racetrack memory with notched geometry},\ }\href
  {https://doi.org/10.1103/PhysRevApplied.17.064019} {\bibfield  {journal}
  {\bibinfo  {journal} {Phys. Rev. Appl.}\ }\textbf {\bibinfo {volume} {17}},\
  \bibinfo {pages} {064019} (\bibinfo {year} {2022})}\BibitemShut {NoStop}%
\bibitem [{\citenamefont {He}\ \emph {et~al.}(2023)\citenamefont {He},
  \citenamefont {Tomasello}, \citenamefont {Luo}, \citenamefont {Zhang},
  \citenamefont {Nie}, \citenamefont {Carpentieri}, \citenamefont {Han},
  \citenamefont {Finocchio},\ and\ \citenamefont {Yu}}]{he2023all}%
  \BibitemOpen
  \bibfield  {author} {\bibinfo {author} {\bibfnamefont {B.}~\bibnamefont
  {He}}, \bibinfo {author} {\bibfnamefont {R.}~\bibnamefont {Tomasello}},
  \bibinfo {author} {\bibfnamefont {X.}~\bibnamefont {Luo}}, \bibinfo {author}
  {\bibfnamefont {R.}~\bibnamefont {Zhang}}, \bibinfo {author} {\bibfnamefont
  {Z.}~\bibnamefont {Nie}}, \bibinfo {author} {\bibfnamefont {M.}~\bibnamefont
  {Carpentieri}}, \bibinfo {author} {\bibfnamefont {X.}~\bibnamefont {Han}},
  \bibinfo {author} {\bibfnamefont {G.}~\bibnamefont {Finocchio}},\ and\
  \bibinfo {author} {\bibfnamefont {G.}~\bibnamefont {Yu}},\ }\bibfield
  {title} {\bibinfo {title} {All-electrical 9-bit skyrmion-based racetrack
  memory designed with laser irradiation},\ }\href
  {https://doi.org/https://pubs.acs.org/doi/abs/10.1021/acs.nanolett.3c02978}
  {\bibfield  {journal} {\bibinfo  {journal} {Nano Lett.}\ }\textbf {\bibinfo
  {volume} {23}},\ \bibinfo {pages} {9482} (\bibinfo {year}
  {2023})}\BibitemShut {NoStop}%
\bibitem [{\citenamefont {Dzyaloshinsky}(1958)}]{DZYALOSHINSKY1958241}%
  \BibitemOpen
  \bibfield  {author} {\bibinfo {author} {\bibfnamefont {I.}~\bibnamefont
  {Dzyaloshinsky}},\ }\bibfield  {title} {\bibinfo {title} {A thermodynamic
  theory of “weak” ferromagnetism of antiferromagnetics},\ }\href
  {https://doi.org/https://doi.org/10.1016/0022-3697(58)90076-3} {\bibfield
  {journal} {\bibinfo  {journal} {J. Phys. Chem. Solids}\ }\textbf {\bibinfo
  {volume} {4}},\ \bibinfo {pages} {241} (\bibinfo {year} {1958})}\BibitemShut
  {NoStop}%
\bibitem [{\citenamefont {Moriya}(1960)}]{PhysRev.120.91}%
  \BibitemOpen
  \bibfield  {author} {\bibinfo {author} {\bibfnamefont {T.}~\bibnamefont
  {Moriya}},\ }\bibfield  {title} {\bibinfo {title} {Anisotropic
  {S}uperexchange {I}nteraction and {W}eak {F}erromagnetism},\ }\href
  {https://doi.org/10.1103/PhysRev.120.91} {\bibfield  {journal} {\bibinfo
  {journal} {Phys. Rev.}\ }\textbf {\bibinfo {volume} {120}},\ \bibinfo {pages}
  {91} (\bibinfo {year} {1960})}\BibitemShut {NoStop}%
\bibitem [{\citenamefont {Mermin}\ and\ \citenamefont
  {Wagner}(1966)}]{mermin1966absence}%
  \BibitemOpen
  \bibfield  {author} {\bibinfo {author} {\bibfnamefont {N.~D.}\ \bibnamefont
  {Mermin}}\ and\ \bibinfo {author} {\bibfnamefont {H.}~\bibnamefont
  {Wagner}},\ }\bibfield  {title} {\bibinfo {title} {Absence of ferromagnetism
  or antiferromagnetism in one-or two-dimensional isotropic {H}eisenberg
  models},\ }\href
  {https://doi.org/https://doi.org/10.1103/PhysRevLett.17.1133} {\bibfield
  {journal} {\bibinfo  {journal} {Phys. Rev. Lett.}\ }\textbf {\bibinfo
  {volume} {17}},\ \bibinfo {pages} {1133} (\bibinfo {year}
  {1966})}\BibitemShut {NoStop}%
\bibitem [{\citenamefont {Bander}\ and\ \citenamefont
  {Mills}(1988)}]{bander1988ferromagnetism}%
  \BibitemOpen
  \bibfield  {author} {\bibinfo {author} {\bibfnamefont {M.}~\bibnamefont
  {Bander}}\ and\ \bibinfo {author} {\bibfnamefont {D.}~\bibnamefont {Mills}},\
  }\bibfield  {title} {\bibinfo {title} {Ferromagnetism of ultrathin films},\
  }\href {https://doi.org/https://doi.org/10.1103/PhysRevB.38.12015} {\bibfield
   {journal} {\bibinfo  {journal} {Phys. Rev. B}\ }\textbf {\bibinfo {volume}
  {38}},\ \bibinfo {pages} {12015} (\bibinfo {year} {1988})}\BibitemShut
  {NoStop}%
\bibitem [{\citenamefont {Wilson}\ \emph {et~al.}(2014)\citenamefont {Wilson},
  \citenamefont {Butenko}, \citenamefont {Bogdanov},\ and\ \citenamefont
  {Monchesky}}]{wilson2014chiral}%
  \BibitemOpen
  \bibfield  {author} {\bibinfo {author} {\bibfnamefont {M.~N.}\ \bibnamefont
  {Wilson}}, \bibinfo {author} {\bibfnamefont {A.}~\bibnamefont {Butenko}},
  \bibinfo {author} {\bibfnamefont {A.}~\bibnamefont {Bogdanov}},\ and\
  \bibinfo {author} {\bibfnamefont {T.}~\bibnamefont {Monchesky}},\ }\bibfield
  {title} {\bibinfo {title} {Chiral skyrmions in cubic helimagnet films: {T}he
  role of uniaxial anisotropy},\ }\href
  {https://doi.org/https://doi.org/10.1103/PhysRevB.89.094411} {\bibfield
  {journal} {\bibinfo  {journal} {Phys. Rev. B}\ }\textbf {\bibinfo {volume}
  {89}},\ \bibinfo {pages} {094411} (\bibinfo {year} {2014})}\BibitemShut
  {NoStop}%
\bibitem [{\citenamefont {Bogdanov}\ and\ \citenamefont
  {Hubert}(1994)}]{bogdanov1994thermodynamically}%
  \BibitemOpen
  \bibfield  {author} {\bibinfo {author} {\bibfnamefont {A.}~\bibnamefont
  {Bogdanov}}\ and\ \bibinfo {author} {\bibfnamefont {A.}~\bibnamefont
  {Hubert}},\ }\bibfield  {title} {\bibinfo {title} {Thermodynamically stable
  magnetic vortex states in magnetic crystals},\ }\href
  {https://doi.org/https://doi.org/10.1016/0304-8853(94)90046-9} {\bibfield
  {journal} {\bibinfo  {journal} {J. Magn. Magn. Mater.}\ }\textbf {\bibinfo
  {volume} {138}},\ \bibinfo {pages} {255} (\bibinfo {year}
  {1994})}\BibitemShut {NoStop}%
\bibitem [{\citenamefont {Zhang}\ \emph {et~al.}(2018)\citenamefont {Zhang},
  \citenamefont {Wang}, \citenamefont {Burn}, \citenamefont {Peng},
  \citenamefont {Berger}, \citenamefont {Bauer}, \citenamefont {Pfleiderer},
  \citenamefont {Van Der~Laan},\ and\ \citenamefont
  {Hesjedal}}]{zhang2018manipulation}%
  \BibitemOpen
  \bibfield  {author} {\bibinfo {author} {\bibfnamefont {S.}~\bibnamefont
  {Zhang}}, \bibinfo {author} {\bibfnamefont {W.}~\bibnamefont {Wang}},
  \bibinfo {author} {\bibfnamefont {D.}~\bibnamefont {Burn}}, \bibinfo {author}
  {\bibfnamefont {H.}~\bibnamefont {Peng}}, \bibinfo {author} {\bibfnamefont
  {H.}~\bibnamefont {Berger}}, \bibinfo {author} {\bibfnamefont
  {A.}~\bibnamefont {Bauer}}, \bibinfo {author} {\bibfnamefont
  {C.}~\bibnamefont {Pfleiderer}}, \bibinfo {author} {\bibfnamefont
  {G.}~\bibnamefont {Van Der~Laan}},\ and\ \bibinfo {author} {\bibfnamefont
  {T.}~\bibnamefont {Hesjedal}},\ }\bibfield  {title} {\bibinfo {title}
  {Manipulation of skyrmion motion by magnetic field gradients},\ }\href
  {https://doi.org/https://doi.org/10.1038/s41467-018-04563-4} {\bibfield
  {journal} {\bibinfo  {journal} {Nat. Commun.}\ }\textbf {\bibinfo {volume}
  {9}},\ \bibinfo {pages} {2115} (\bibinfo {year} {2018})}\BibitemShut
  {NoStop}%
\bibitem [{\citenamefont {M\"uhlbauer}\ \emph {et~al.}(2009)\citenamefont
  {M\"uhlbauer}, \citenamefont {Binz}, \citenamefont {Jonietz}, \citenamefont
  {Pfleiderer}, \citenamefont {Rosch}, \citenamefont {Neubauer}, \citenamefont
  {Georgii},\ and\ \citenamefont {Boni}}]{Muhlbauer2009skyrmion}%
  \BibitemOpen
  \bibfield  {author} {\bibinfo {author} {\bibfnamefont {S.}~\bibnamefont
  {M\"uhlbauer}}, \bibinfo {author} {\bibfnamefont {B.}~\bibnamefont {Binz}},
  \bibinfo {author} {\bibfnamefont {F.}~\bibnamefont {Jonietz}}, \bibinfo
  {author} {\bibfnamefont {C.}~\bibnamefont {Pfleiderer}}, \bibinfo {author}
  {\bibfnamefont {A.}~\bibnamefont {Rosch}}, \bibinfo {author} {\bibfnamefont
  {A.}~\bibnamefont {Neubauer}}, \bibinfo {author} {\bibfnamefont
  {R.}~\bibnamefont {Georgii}},\ and\ \bibinfo {author} {\bibfnamefont
  {P.}~\bibnamefont {Boni}},\ }\bibfield  {title} {\bibinfo {title} {Skyrmion
  lattice in a chiral magnet},\ }\href
  {https://doi.org/10.1126/science.1166767} {\bibfield  {journal} {\bibinfo
  {journal} {Science}\ }\textbf {\bibinfo {volume} {323}},\ \bibinfo {pages}
  {915} (\bibinfo {year} {2009})}\BibitemShut {NoStop}%
\bibitem [{\citenamefont {Yu}\ \emph {et~al.}(2010)\citenamefont {Yu},
  \citenamefont {Onose}, \citenamefont {Kanazawa}, \citenamefont {Park},
  \citenamefont {Han}, \citenamefont {Matsui}, \citenamefont {Nagaosa},\ and\
  \citenamefont {Tokura}}]{yu2010real}%
  \BibitemOpen
  \bibfield  {author} {\bibinfo {author} {\bibfnamefont {X.}~\bibnamefont
  {Yu}}, \bibinfo {author} {\bibfnamefont {Y.}~\bibnamefont {Onose}}, \bibinfo
  {author} {\bibfnamefont {N.}~\bibnamefont {Kanazawa}}, \bibinfo {author}
  {\bibfnamefont {J.~H.}\ \bibnamefont {Park}}, \bibinfo {author}
  {\bibfnamefont {J.}~\bibnamefont {Han}}, \bibinfo {author} {\bibfnamefont
  {Y.}~\bibnamefont {Matsui}}, \bibinfo {author} {\bibfnamefont
  {N.}~\bibnamefont {Nagaosa}},\ and\ \bibinfo {author} {\bibfnamefont
  {Y.}~\bibnamefont {Tokura}},\ }\bibfield  {title} {\bibinfo {title}
  {Real-space observation of a two-dimensional skyrmion crystal},\ }\href
  {https://doi.org/10.1038/nature09124} {\bibfield  {journal} {\bibinfo
  {journal} {Nature}\ }\textbf {\bibinfo {volume} {465}},\ \bibinfo {pages}
  {901} (\bibinfo {year} {2010})}\BibitemShut {NoStop}%
\bibitem [{\citenamefont {Yu}\ \emph {et~al.}(2011)\citenamefont {Yu},
  \citenamefont {Kanazawa}, \citenamefont {Onose}, \citenamefont {Kimoto},
  \citenamefont {Zhang}, \citenamefont {Ishiwata}, \citenamefont {Matsui},\
  and\ \citenamefont {Tokura}}]{yu2011near}%
  \BibitemOpen
  \bibfield  {author} {\bibinfo {author} {\bibfnamefont {X.}~\bibnamefont
  {Yu}}, \bibinfo {author} {\bibfnamefont {N.}~\bibnamefont {Kanazawa}},
  \bibinfo {author} {\bibfnamefont {Y.}~\bibnamefont {Onose}}, \bibinfo
  {author} {\bibfnamefont {K.}~\bibnamefont {Kimoto}}, \bibinfo {author}
  {\bibfnamefont {W.}~\bibnamefont {Zhang}}, \bibinfo {author} {\bibfnamefont
  {S.}~\bibnamefont {Ishiwata}}, \bibinfo {author} {\bibfnamefont
  {Y.}~\bibnamefont {Matsui}},\ and\ \bibinfo {author} {\bibfnamefont
  {Y.}~\bibnamefont {Tokura}},\ }\bibfield  {title} {\bibinfo {title} {Near
  room-temperature formation of a skyrmion crystal in thin-films of the
  helimagnet {F}e{G}e},\ }\href
  {https://doi.org/https://doi.org/10.1038/nmat2916} {\bibfield  {journal}
  {\bibinfo  {journal} {Nat. Mater.}\ }\textbf {\bibinfo {volume} {10}},\
  \bibinfo {pages} {106} (\bibinfo {year} {2011})}\BibitemShut {NoStop}%
\bibitem [{\citenamefont {Yu}\ \emph {et~al.}(2012)\citenamefont {Yu},
  \citenamefont {Mostovoy}, \citenamefont {Tokunaga}, \citenamefont {Zhang},
  \citenamefont {Kimoto}, \citenamefont {Matsui}, \citenamefont {Kaneko},
  \citenamefont {Nagaosa},\ and\ \citenamefont {Tokura}}]{yu2012magnetic}%
  \BibitemOpen
  \bibfield  {author} {\bibinfo {author} {\bibfnamefont {X.}~\bibnamefont
  {Yu}}, \bibinfo {author} {\bibfnamefont {M.}~\bibnamefont {Mostovoy}},
  \bibinfo {author} {\bibfnamefont {Y.}~\bibnamefont {Tokunaga}}, \bibinfo
  {author} {\bibfnamefont {W.}~\bibnamefont {Zhang}}, \bibinfo {author}
  {\bibfnamefont {K.}~\bibnamefont {Kimoto}}, \bibinfo {author} {\bibfnamefont
  {Y.}~\bibnamefont {Matsui}}, \bibinfo {author} {\bibfnamefont
  {Y.}~\bibnamefont {Kaneko}}, \bibinfo {author} {\bibfnamefont
  {N.}~\bibnamefont {Nagaosa}},\ and\ \bibinfo {author} {\bibfnamefont
  {Y.}~\bibnamefont {Tokura}},\ }\bibfield  {title} {\bibinfo {title} {Magnetic
  stripes and skyrmions with helicity reversals},\ }\href
  {https://doi.org/https://doi.org/10.1073/pnas.1118496109} {\bibfield
  {journal} {\bibinfo  {journal} {Proc. Natl. Acad. Sci.}\ }\textbf {\bibinfo
  {volume} {109}},\ \bibinfo {pages} {8856} (\bibinfo {year}
  {2012})}\BibitemShut {NoStop}%
\bibitem [{\citenamefont {Woo}\ \emph {et~al.}(2016)\citenamefont {Woo},
  \citenamefont {Litzius}, \citenamefont {Kr{\"u}ger}, \citenamefont {Im},
  \citenamefont {Caretta}, \citenamefont {Richter}, \citenamefont {Mann},
  \citenamefont {Krone}, \citenamefont {Reeve}, \citenamefont {Weigand} \emph
  {et~al.}}]{woo2016observation}%
  \BibitemOpen
  \bibfield  {author} {\bibinfo {author} {\bibfnamefont {S.}~\bibnamefont
  {Woo}}, \bibinfo {author} {\bibfnamefont {K.}~\bibnamefont {Litzius}},
  \bibinfo {author} {\bibfnamefont {B.}~\bibnamefont {Kr{\"u}ger}}, \bibinfo
  {author} {\bibfnamefont {M.-Y.}\ \bibnamefont {Im}}, \bibinfo {author}
  {\bibfnamefont {L.}~\bibnamefont {Caretta}}, \bibinfo {author} {\bibfnamefont
  {K.}~\bibnamefont {Richter}}, \bibinfo {author} {\bibfnamefont
  {M.}~\bibnamefont {Mann}}, \bibinfo {author} {\bibfnamefont {A.}~\bibnamefont
  {Krone}}, \bibinfo {author} {\bibfnamefont {R.~M.}\ \bibnamefont {Reeve}},
  \bibinfo {author} {\bibfnamefont {M.}~\bibnamefont {Weigand}}, \emph
  {et~al.},\ }\bibfield  {title} {\bibinfo {title} {Observation of
  room-temperature magnetic skyrmions and their current-driven dynamics in
  ultrathin metallic ferromagnets},\ }\href
  {https://doi.org/https://doi.org/10.1038/nmat4593} {\bibfield  {journal}
  {\bibinfo  {journal} {Nat. Mater.}\ }\textbf {\bibinfo {volume} {15}},\
  \bibinfo {pages} {501} (\bibinfo {year} {2016})}\BibitemShut {NoStop}%
\bibitem [{\citenamefont {Cort{\'e}s-Ortu{\~n}o}\ \emph
  {et~al.}(2019)\citenamefont {Cort{\'e}s-Ortu{\~n}o}, \citenamefont {Romming},
  \citenamefont {Beg}, \citenamefont {Von~Bergmann}, \citenamefont {Kubetzka},
  \citenamefont {Hovorka}, \citenamefont {Fangohr},\ and\ \citenamefont
  {Wiesendanger}}]{cortes2019nanoscale}%
  \BibitemOpen
  \bibfield  {author} {\bibinfo {author} {\bibfnamefont {D.}~\bibnamefont
  {Cort{\'e}s-Ortu{\~n}o}}, \bibinfo {author} {\bibfnamefont {N.}~\bibnamefont
  {Romming}}, \bibinfo {author} {\bibfnamefont {M.}~\bibnamefont {Beg}},
  \bibinfo {author} {\bibfnamefont {K.}~\bibnamefont {Von~Bergmann}}, \bibinfo
  {author} {\bibfnamefont {A.}~\bibnamefont {Kubetzka}}, \bibinfo {author}
  {\bibfnamefont {O.}~\bibnamefont {Hovorka}}, \bibinfo {author} {\bibfnamefont
  {H.}~\bibnamefont {Fangohr}},\ and\ \bibinfo {author} {\bibfnamefont
  {R.}~\bibnamefont {Wiesendanger}},\ }\bibfield  {title} {\bibinfo {title}
  {Nanoscale magnetic skyrmions and target states in confined geometries},\
  }\href {https://doi.org/https://doi.org/10.1103/PhysRevB.99.214408}
  {\bibfield  {journal} {\bibinfo  {journal} {Phys. Rev. B}\ }\textbf {\bibinfo
  {volume} {99}},\ \bibinfo {pages} {214408} (\bibinfo {year}
  {2019})}\BibitemShut {NoStop}%
\bibitem [{\citenamefont {Zhang}\ \emph {et~al.}(2023)\citenamefont {Zhang},
  \citenamefont {Wang}, \citenamefont {Zhang}, \citenamefont {Huang},
  \citenamefont {Zhu}, \citenamefont {Liu}, \citenamefont {Zhang},
  \citenamefont {Han}, \citenamefont {Yang}, \citenamefont {Zhang} \emph
  {et~al.}}]{zhang2023direct}%
  \BibitemOpen
  \bibfield  {author} {\bibinfo {author} {\bibfnamefont {J.}~\bibnamefont
  {Zhang}}, \bibinfo {author} {\bibfnamefont {M.}~\bibnamefont {Wang}},
  \bibinfo {author} {\bibfnamefont {Z.}~\bibnamefont {Zhang}}, \bibinfo
  {author} {\bibfnamefont {H.}~\bibnamefont {Huang}}, \bibinfo {author}
  {\bibfnamefont {L.}~\bibnamefont {Zhu}}, \bibinfo {author} {\bibfnamefont
  {D.}~\bibnamefont {Liu}}, \bibinfo {author} {\bibfnamefont {H.}~\bibnamefont
  {Zhang}}, \bibinfo {author} {\bibfnamefont {F.}~\bibnamefont {Han}}, \bibinfo
  {author} {\bibfnamefont {H.}~\bibnamefont {Yang}}, \bibinfo {author}
  {\bibfnamefont {J.}~\bibnamefont {Zhang}}, \emph {et~al.},\ }\bibfield
  {title} {\bibinfo {title} {Direct observation of magnetic skyrmions and their
  current-induced dynamics in epitaxial single-crystal oxide films},\ }\href
  {https://doi.org/https://pubs.acs.org/doi/abs/10.1021/acs.nanolett.3c00342}
  {\bibfield  {journal} {\bibinfo  {journal} {Nano Lett.}\ }\textbf {\bibinfo
  {volume} {23}},\ \bibinfo {pages} {4258} (\bibinfo {year}
  {2023})}\BibitemShut {NoStop}%
\bibitem [{\citenamefont {Ding}\ \emph {et~al.}(2019)\citenamefont {Ding},
  \citenamefont {Li}, \citenamefont {Xu}, \citenamefont {Li}, \citenamefont
  {Hou}, \citenamefont {Liu}, \citenamefont {Xi}, \citenamefont {Xu},
  \citenamefont {Yao},\ and\ \citenamefont {Wang}}]{ding2019observation}%
  \BibitemOpen
  \bibfield  {author} {\bibinfo {author} {\bibfnamefont {B.}~\bibnamefont
  {Ding}}, \bibinfo {author} {\bibfnamefont {Z.}~\bibnamefont {Li}}, \bibinfo
  {author} {\bibfnamefont {G.}~\bibnamefont {Xu}}, \bibinfo {author}
  {\bibfnamefont {H.}~\bibnamefont {Li}}, \bibinfo {author} {\bibfnamefont
  {Z.}~\bibnamefont {Hou}}, \bibinfo {author} {\bibfnamefont {E.}~\bibnamefont
  {Liu}}, \bibinfo {author} {\bibfnamefont {X.}~\bibnamefont {Xi}}, \bibinfo
  {author} {\bibfnamefont {F.}~\bibnamefont {Xu}}, \bibinfo {author}
  {\bibfnamefont {Y.}~\bibnamefont {Yao}},\ and\ \bibinfo {author}
  {\bibfnamefont {W.}~\bibnamefont {Wang}},\ }\bibfield  {title} {\bibinfo
  {title} {Observation of magnetic skyrmion bubbles in a van der waals
  ferromagnet {Fe$_3$GeTe$_2$}},\ }\href
  {https://doi.org/10.1021/acs.nanolett.9b03453} {\bibfield  {journal}
  {\bibinfo  {journal} {Nano Lett.}\ }\textbf {\bibinfo {volume} {20}},\
  \bibinfo {pages} {868} (\bibinfo {year} {2019})}\BibitemShut {NoStop}%
\bibitem [{\citenamefont {Zhang}\ \emph {et~al.}(2022)\citenamefont {Zhang},
  \citenamefont {Raftrey}, \citenamefont {Chan}, \citenamefont {Shao},
  \citenamefont {Chen}, \citenamefont {Chen}, \citenamefont {Huang},
  \citenamefont {Reichanadter}, \citenamefont {Dong}, \citenamefont {Susarla}
  \emph {et~al.}}]{zhang2022room}%
  \BibitemOpen
  \bibfield  {author} {\bibinfo {author} {\bibfnamefont {H.}~\bibnamefont
  {Zhang}}, \bibinfo {author} {\bibfnamefont {D.}~\bibnamefont {Raftrey}},
  \bibinfo {author} {\bibfnamefont {Y.-T.}\ \bibnamefont {Chan}}, \bibinfo
  {author} {\bibfnamefont {Y.-T.}\ \bibnamefont {Shao}}, \bibinfo {author}
  {\bibfnamefont {R.}~\bibnamefont {Chen}}, \bibinfo {author} {\bibfnamefont
  {X.}~\bibnamefont {Chen}}, \bibinfo {author} {\bibfnamefont {X.}~\bibnamefont
  {Huang}}, \bibinfo {author} {\bibfnamefont {J.~T.}\ \bibnamefont
  {Reichanadter}}, \bibinfo {author} {\bibfnamefont {K.}~\bibnamefont {Dong}},
  \bibinfo {author} {\bibfnamefont {S.}~\bibnamefont {Susarla}}, \emph
  {et~al.},\ }\bibfield  {title} {\bibinfo {title} {Room-temperature skyrmion
  lattice in a layered magnet {(Fe$_{0. 5}$Co$_{0. 5}$)$_5$GeTe$_2$}},\ }\href
  {https://doi.org/10.1126/sciadv.abm7103} {\bibfield  {journal} {\bibinfo
  {journal} {Sci. Adv.}\ }\textbf {\bibinfo {volume} {8}},\ \bibinfo {pages}
  {eabm7103} (\bibinfo {year} {2022})}\BibitemShut {NoStop}%
\bibitem [{\citenamefont {Roessler}\ \emph {et~al.}(2006)\citenamefont
  {Roessler}, \citenamefont {Bogdanov},\ and\ \citenamefont
  {Pfleiderer}}]{roessler2006spontaneous}%
  \BibitemOpen
  \bibfield  {author} {\bibinfo {author} {\bibfnamefont {U.~K.}\ \bibnamefont
  {Roessler}}, \bibinfo {author} {\bibfnamefont {A.}~\bibnamefont {Bogdanov}},\
  and\ \bibinfo {author} {\bibfnamefont {C.}~\bibnamefont {Pfleiderer}},\
  }\bibfield  {title} {\bibinfo {title} {Spontaneous skyrmion ground states in
  magnetic metals},\ }\href
  {https://doi.org/https://doi.org/10.1038/nature05056} {\bibfield  {journal}
  {\bibinfo  {journal} {Nature}\ }\textbf {\bibinfo {volume} {442}},\ \bibinfo
  {pages} {797} (\bibinfo {year} {2006})}\BibitemShut {NoStop}%
\bibitem [{\citenamefont {Leonov}\ \emph {et~al.}(2016)\citenamefont {Leonov},
  \citenamefont {Monchesky}, \citenamefont {Romming}, \citenamefont {Kubetzka},
  \citenamefont {Bogdanov},\ and\ \citenamefont
  {Wiesendanger}}]{leonov2016properties}%
  \BibitemOpen
  \bibfield  {author} {\bibinfo {author} {\bibfnamefont {A.}~\bibnamefont
  {Leonov}}, \bibinfo {author} {\bibfnamefont {T.}~\bibnamefont {Monchesky}},
  \bibinfo {author} {\bibfnamefont {N.}~\bibnamefont {Romming}}, \bibinfo
  {author} {\bibfnamefont {A.}~\bibnamefont {Kubetzka}}, \bibinfo {author}
  {\bibfnamefont {A.}~\bibnamefont {Bogdanov}},\ and\ \bibinfo {author}
  {\bibfnamefont {R.}~\bibnamefont {Wiesendanger}},\ }\bibfield  {title}
  {\bibinfo {title} {The properties of isolated chiral skyrmions in thin
  magnetic films},\ }\href
  {https://doi.org/https://doi.org/10.1103/PhysRevApplied.17.064019} {\bibfield
   {journal} {\bibinfo  {journal} {New J. Phys.}\ }\textbf {\bibinfo {volume}
  {18}},\ \bibinfo {pages} {065003} (\bibinfo {year} {2016})}\BibitemShut
  {NoStop}%
\bibitem [{\citenamefont {Wijethunga}\ \emph {et~al.}(2025)\citenamefont
  {Wijethunga}, \citenamefont {Hu},\ and\ \citenamefont
  {Wang}}]{wijethunga2025phase}%
  \BibitemOpen
  \bibfield  {author} {\bibinfo {author} {\bibfnamefont {M.~V.}\ \bibnamefont
  {Wijethunga}}, \bibinfo {author} {\bibfnamefont {X.}~\bibnamefont {Hu}},\
  and\ \bibinfo {author} {\bibfnamefont {X.}~\bibnamefont {Wang}},\ }\bibfield
  {title} {\bibinfo {title} {Phase-diagram of condensed and isolated skyrmions
  in chiral magnetic films},\ }\href
  {https://doi.org/10.1038/s42005-025-01980-y} {\bibfield  {journal} {\bibinfo
  {journal} {Commun. Phys.}\ }\textbf {\bibinfo {volume} {8}},\ \bibinfo
  {pages} {123} (\bibinfo {year} {2025})}\BibitemShut {NoStop}%
\bibitem [{\citenamefont {Wang}\ \emph {et~al.}(2018)\citenamefont {Wang},
  \citenamefont {Yuan},\ and\ \citenamefont {Wang}}]{wang2018theory}%
  \BibitemOpen
  \bibfield  {author} {\bibinfo {author} {\bibfnamefont {X.}~\bibnamefont
  {Wang}}, \bibinfo {author} {\bibfnamefont {H.}~\bibnamefont {Yuan}},\ and\
  \bibinfo {author} {\bibfnamefont {X.}~\bibnamefont {Wang}},\ }\bibfield
  {title} {\bibinfo {title} {A theory on skyrmion size},\ }\href
  {https://doi.org/https://doi.org/10.1038/s42005-018-0029-0} {\bibfield
  {journal} {\bibinfo  {journal} {Commun. Phys.}\ }\textbf {\bibinfo {volume}
  {1}},\ \bibinfo {pages} {31} (\bibinfo {year} {2018})}\BibitemShut {NoStop}%
\bibitem [{\citenamefont {Wu}\ \emph {et~al.}(2021)\citenamefont {Wu},
  \citenamefont {Hu}, \citenamefont {Jing},\ and\ \citenamefont
  {Wang}}]{wu2021size}%
  \BibitemOpen
  \bibfield  {author} {\bibinfo {author} {\bibfnamefont {H.}~\bibnamefont
  {Wu}}, \bibinfo {author} {\bibfnamefont {X.}~\bibnamefont {Hu}}, \bibinfo
  {author} {\bibfnamefont {K.}~\bibnamefont {Jing}},\ and\ \bibinfo {author}
  {\bibfnamefont {X.}~\bibnamefont {Wang}},\ }\bibfield  {title} {\bibinfo
  {title} {Size and profile of skyrmions in skyrmion crystals},\ }\href
  {https://doi.org/10.1038/s42005-021-00716-y} {\bibfield  {journal} {\bibinfo
  {journal} {Commun. Phys.}\ }\textbf {\bibinfo {volume} {4}},\ \bibinfo
  {pages} {210} (\bibinfo {year} {2021})}\BibitemShut {NoStop}%
\bibitem [{\citenamefont {Hu}\ \emph {et~al.}(2022)\citenamefont {Hu},
  \citenamefont {Wu},\ and\ \citenamefont {Wang}}]{hu2022theory}%
  \BibitemOpen
  \bibfield  {author} {\bibinfo {author} {\bibfnamefont {X.-C.}\ \bibnamefont
  {Hu}}, \bibinfo {author} {\bibfnamefont {H.-T.}\ \bibnamefont {Wu}},\ and\
  \bibinfo {author} {\bibfnamefont {X.}~\bibnamefont {Wang}},\ }\bibfield
  {title} {\bibinfo {title} {A theory of skyrmion crystal formation},\ }\href
  {https://doi.org/10.1039/D2NR01300B} {\bibfield  {journal} {\bibinfo
  {journal} {Nanoscale}\ }\textbf {\bibinfo {volume} {14}},\ \bibinfo {pages}
  {7516} (\bibinfo {year} {2022})}\BibitemShut {NoStop}%
\bibitem [{\citenamefont {Wu}\ \emph {et~al.}(2022)\citenamefont {Wu},
  \citenamefont {Hu},\ and\ \citenamefont {Wang}}]{wu2022nematic}%
  \BibitemOpen
  \bibfield  {author} {\bibinfo {author} {\bibfnamefont {H.-T.}\ \bibnamefont
  {Wu}}, \bibinfo {author} {\bibfnamefont {X.-C.}\ \bibnamefont {Hu}},\ and\
  \bibinfo {author} {\bibfnamefont {X.}~\bibnamefont {Wang}},\ }\bibfield
  {title} {\bibinfo {title} {Nematic and smectic stripe phases and stripe-{SkX}
  transformations},\ }\href {https://doi.org/10.1007/s11433-021-1852-8}
  {\bibfield  {journal} {\bibinfo  {journal} {Sci. China:Phys., Mech. Astron.}\
  }\textbf {\bibinfo {volume} {65}},\ \bibinfo {pages} {247512} (\bibinfo
  {year} {2022})}\BibitemShut {NoStop}%
\bibitem [{\citenamefont {Feng}\ \emph {et~al.}(2024)\citenamefont {Feng},
  \citenamefont {Guan}, \citenamefont {Wu}, \citenamefont {Wu},\ and\
  \citenamefont {Song}}]{feng2024classification}%
  \BibitemOpen
  \bibfield  {author} {\bibinfo {author} {\bibfnamefont {D.}~\bibnamefont
  {Feng}}, \bibinfo {author} {\bibfnamefont {Z.}~\bibnamefont {Guan}}, \bibinfo
  {author} {\bibfnamefont {X.}~\bibnamefont {Wu}}, \bibinfo {author}
  {\bibfnamefont {Y.}~\bibnamefont {Wu}},\ and\ \bibinfo {author}
  {\bibfnamefont {C.}~\bibnamefont {Song}},\ }\bibfield  {title} {\bibinfo
  {title} {Classification of skyrmionic textures and extraction of
  {H}amiltonian parameters via machine learning},\ }\href
  {https://doi.org/10.1103/PhysRevApplied.21.034009} {\bibfield  {journal}
  {\bibinfo  {journal} {Phys. Rev. Appl.}\ }\textbf {\bibinfo {volume} {21}},\
  \bibinfo {pages} {034009} (\bibinfo {year} {2024})}\BibitemShut {NoStop}%
\bibitem [{\citenamefont {Hu}\ \emph {et~al.}(2025)\citenamefont {Hu},
  \citenamefont {Chen}, \citenamefont {Zhu}, \citenamefont {Guan},
  \citenamefont {Wu},\ and\ \citenamefont {Song}}]{hu2025machine}%
  \BibitemOpen
  \bibfield  {author} {\bibinfo {author} {\bibfnamefont {H.}~\bibnamefont
  {Hu}}, \bibinfo {author} {\bibfnamefont {Z.}~\bibnamefont {Chen}}, \bibinfo
  {author} {\bibfnamefont {S.}~\bibnamefont {Zhu}}, \bibinfo {author}
  {\bibfnamefont {X.}~\bibnamefont {Guan}}, \bibinfo {author} {\bibfnamefont
  {X.}~\bibnamefont {Wu}},\ and\ \bibinfo {author} {\bibfnamefont
  {C.}~\bibnamefont {Song}},\ }\bibfield  {title} {\bibinfo {title} {Machine
  learning-driven prediction of skyrmion phase boundaries in 2{D} magnets},\
  }\href {https://doi.org/10.1103/55z5-cm5q} {\bibfield  {journal} {\bibinfo
  {journal} {Phys. Rev. Mater.}\ }\textbf {\bibinfo {volume} {9}},\ \bibinfo
  {pages} {074001} (\bibinfo {year} {2025})}\BibitemShut {NoStop}%
\bibitem [{\citenamefont {Tang}\ \emph {et~al.}(2023)\citenamefont {Tang},
  \citenamefont {Weng},\ and\ \citenamefont {Zhang}}]{tang2023neural}%
  \BibitemOpen
  \bibfield  {author} {\bibinfo {author} {\bibfnamefont {Y.}~\bibnamefont
  {Tang}}, \bibinfo {author} {\bibfnamefont {J.}~\bibnamefont {Weng}},\ and\
  \bibinfo {author} {\bibfnamefont {P.}~\bibnamefont {Zhang}},\ }\bibfield
  {title} {\bibinfo {title} {Neural-network solutions to stochastic reaction
  networks},\ }\href {https://doi.org/10.1038/s42256-023-00632-6} {\bibfield
  {journal} {\bibinfo  {journal} {Nat. Mach. Intell.}\ }\textbf {\bibinfo
  {volume} {5}},\ \bibinfo {pages} {376} (\bibinfo {year} {2023})}\BibitemShut
  {NoStop}%
\bibitem [{\citenamefont {Li}\ \emph {et~al.}(2024)\citenamefont {Li},
  \citenamefont {Tang}, \citenamefont {Chen}, \citenamefont {Sun},
  \citenamefont {Zhao}, \citenamefont {Li}, \citenamefont {Tao}, \citenamefont
  {Yuan}, \citenamefont {Duan},\ and\ \citenamefont {Xu}}]{li2024neural}%
  \BibitemOpen
  \bibfield  {author} {\bibinfo {author} {\bibfnamefont {Y.}~\bibnamefont
  {Li}}, \bibinfo {author} {\bibfnamefont {Z.}~\bibnamefont {Tang}}, \bibinfo
  {author} {\bibfnamefont {Z.}~\bibnamefont {Chen}}, \bibinfo {author}
  {\bibfnamefont {M.}~\bibnamefont {Sun}}, \bibinfo {author} {\bibfnamefont
  {B.}~\bibnamefont {Zhao}}, \bibinfo {author} {\bibfnamefont {H.}~\bibnamefont
  {Li}}, \bibinfo {author} {\bibfnamefont {H.}~\bibnamefont {Tao}}, \bibinfo
  {author} {\bibfnamefont {Z.}~\bibnamefont {Yuan}}, \bibinfo {author}
  {\bibfnamefont {W.}~\bibnamefont {Duan}},\ and\ \bibinfo {author}
  {\bibfnamefont {Y.}~\bibnamefont {Xu}},\ }\bibfield  {title} {\bibinfo
  {title} {Neural-network density functional theory based on variational energy
  minimization},\ }\href {https://doi.org/10.1103/PhysRevLett.133.076401}
  {\bibfield  {journal} {\bibinfo  {journal} {Phys. Rev. Lett.}\ }\textbf
  {\bibinfo {volume} {133}},\ \bibinfo {pages} {076401} (\bibinfo {year}
  {2024})}\BibitemShut {NoStop}%
\bibitem [{\citenamefont {Carrasquilla}\ and\ \citenamefont
  {Melko}(2017)}]{carrasquilla2017machine}%
  \BibitemOpen
  \bibfield  {author} {\bibinfo {author} {\bibfnamefont {J.}~\bibnamefont
  {Carrasquilla}}\ and\ \bibinfo {author} {\bibfnamefont {R.~G.}\ \bibnamefont
  {Melko}},\ }\bibfield  {title} {\bibinfo {title} {Machine learning phases of
  matter},\ }\href {https://doi.org/10.1038/nphys4035} {\bibfield  {journal}
  {\bibinfo  {journal} {Nat. Phys.}\ }\textbf {\bibinfo {volume} {13}},\
  \bibinfo {pages} {431} (\bibinfo {year} {2017})}\BibitemShut {NoStop}%
\bibitem [{\citenamefont {Wang}\ \emph {et~al.}(2021)\citenamefont {Wang},
  \citenamefont {Wang}, \citenamefont {Zhang}, \citenamefont {Sun},\ and\
  \citenamefont {Xia}}]{wang2021learning}%
  \BibitemOpen
  \bibfield  {author} {\bibinfo {author} {\bibfnamefont {W.}~\bibnamefont
  {Wang}}, \bibinfo {author} {\bibfnamefont {Z.}~\bibnamefont {Wang}}, \bibinfo
  {author} {\bibfnamefont {Y.}~\bibnamefont {Zhang}}, \bibinfo {author}
  {\bibfnamefont {B.}~\bibnamefont {Sun}},\ and\ \bibinfo {author}
  {\bibfnamefont {K.}~\bibnamefont {Xia}},\ }\bibfield  {title} {\bibinfo
  {title} {Learning order parameters from videos of skyrmion dynamical phases
  with neural networks},\ }\href
  {https://doi.org/10.1103/PhysRevApplied.16.014005} {\bibfield  {journal}
  {\bibinfo  {journal} {Phys. Rev. Appl.}\ }\textbf {\bibinfo {volume} {16}},\
  \bibinfo {pages} {014005} (\bibinfo {year} {2021})}\BibitemShut {NoStop}%
\bibitem [{\citenamefont {Xiong}\ \emph {et~al.}(2025)\citenamefont {Xiong},
  \citenamefont {Zhou}, \citenamefont {Hu}, \citenamefont {Nian},\ and\
  \citenamefont {Zheng}}]{xiong2025capturing}%
  \BibitemOpen
  \bibfield  {author} {\bibinfo {author} {\bibfnamefont {L.}~\bibnamefont
  {Xiong}}, \bibinfo {author} {\bibfnamefont {N.-J.}\ \bibnamefont {Zhou}},
  \bibinfo {author} {\bibfnamefont {S.-Q.}\ \bibnamefont {Hu}}, \bibinfo
  {author} {\bibfnamefont {L.-L.}\ \bibnamefont {Nian}},\ and\ \bibinfo
  {author} {\bibfnamefont {B.}~\bibnamefont {Zheng}},\ }\bibfield  {title}
  {\bibinfo {title} {Capturing the dynamics of the phase transition of
  skyrmions with a nonstationary machine learning approach},\ }\href
  {https://doi.org/10.1103/PhysRevB.111.184415} {\bibfield  {journal} {\bibinfo
   {journal} {Phys. Rev. B}\ }\textbf {\bibinfo {volume} {111}},\ \bibinfo
  {pages} {184415} (\bibinfo {year} {2025})}\BibitemShut {NoStop}%
\bibitem [{\citenamefont {M\"uller}\ \emph {et~al.}(2019)\citenamefont
  {M\"uller}, \citenamefont {Hoffmann}, \citenamefont {Di\ss{}elkamp},
  \citenamefont {Sch\"urhoff}, \citenamefont {Mavros}, \citenamefont
  {Sallermann}, \citenamefont {Kiselev}, \citenamefont {J\'onsson},\ and\
  \citenamefont {Bl\"ugel}}]{PhysRevB.99.224414}%
  \BibitemOpen
  \bibfield  {author} {\bibinfo {author} {\bibfnamefont {G.~P.}\ \bibnamefont
  {M\"uller}}, \bibinfo {author} {\bibfnamefont {M.}~\bibnamefont {Hoffmann}},
  \bibinfo {author} {\bibfnamefont {C.}~\bibnamefont {Di\ss{}elkamp}}, \bibinfo
  {author} {\bibfnamefont {D.}~\bibnamefont {Sch\"urhoff}}, \bibinfo {author}
  {\bibfnamefont {S.}~\bibnamefont {Mavros}}, \bibinfo {author} {\bibfnamefont
  {M.}~\bibnamefont {Sallermann}}, \bibinfo {author} {\bibfnamefont {N.~S.}\
  \bibnamefont {Kiselev}}, \bibinfo {author} {\bibfnamefont {H.}~\bibnamefont
  {J\'onsson}},\ and\ \bibinfo {author} {\bibfnamefont {S.}~\bibnamefont
  {Bl\"ugel}},\ }\bibfield  {title} {\bibinfo {title} {Spirit: Multifunctional
  framework for atomistic spin simulations},\ }\href
  {https://doi.org/10.1103/PhysRevB.99.224414} {\bibfield  {journal} {\bibinfo
  {journal} {Phys. Rev. B}\ }\textbf {\bibinfo {volume} {99}},\ \bibinfo
  {pages} {224414} (\bibinfo {year} {2019})}\BibitemShut {NoStop}%
\bibitem [{\citenamefont {Landau}\ \emph {et~al.}(1935)\citenamefont {Landau},
  \citenamefont {Lifshitz} \emph {et~al.}}]{landau1992theory}%
  \BibitemOpen
  \bibfield  {author} {\bibinfo {author} {\bibfnamefont {L.}~\bibnamefont
  {Landau}}, \bibinfo {author} {\bibfnamefont {E.}~\bibnamefont {Lifshitz}},
  \emph {et~al.},\ }\bibfield  {title} {\bibinfo {title} {On the theory of the
  dispersion of magnetic permeability in ferromagnetic bodies},\ }\href
  {https://www.cpt.univ-mrs.fr/~verga/pdfs/Landau-1935fk.pdf} {\bibfield
  {journal} {\bibinfo  {journal} {Phys. Z. Sowjetunion}\ }\textbf {\bibinfo
  {volume} {8}},\ \bibinfo {pages} {101} (\bibinfo {year} {1935})}\BibitemShut
  {NoStop}%
\bibitem [{\citenamefont {Gilbert}(2004)}]{1353448}%
  \BibitemOpen
  \bibfield  {author} {\bibinfo {author} {\bibfnamefont {T.}~\bibnamefont
  {Gilbert}},\ }\bibfield  {title} {\bibinfo {title} {A phenomenological theory
  of damping in ferromagnetic materials},\ }\href
  {https://doi.org/10.1109/TMAG.2004.836740} {\bibfield  {journal} {\bibinfo
  {journal} {IEEE Transactions on Magnetics}\ }\textbf {\bibinfo {volume}
  {40}},\ \bibinfo {pages} {3443} (\bibinfo {year} {2004})}\BibitemShut
  {NoStop}%
\bibitem [{sup()}]{supplmat}%
  \BibitemOpen
  \href@noop {} {}\bibinfo {note} {See Supplemental Material at
  http://link.aps.org/supplemental/ for validation tests, energy competition
  during the dynamical process, symbolic regression models, phase prediction by
  neural networks, skyrmion property extraction, and interrelations among
  skyrmion properties.}\BibitemShut {Stop}%
\bibitem [{\citenamefont {Rohart}\ and\ \citenamefont
  {Thiaville}(2013)}]{rohart2013skyrmion}%
  \BibitemOpen
  \bibfield  {author} {\bibinfo {author} {\bibfnamefont {S.}~\bibnamefont
  {Rohart}}\ and\ \bibinfo {author} {\bibfnamefont {A.}~\bibnamefont
  {Thiaville}},\ }\bibfield  {title} {\bibinfo {title} {Skyrmion confinement in
  ultrathin film nanostructures in the presence of {D}zyaloshinskii-{M}oriya
  interaction},\ }\href {https://doi.org/10.1103/PhysRevB.88.184422} {\bibfield
   {journal} {\bibinfo  {journal} {Phys. Rev. B}\ }\textbf {\bibinfo {volume}
  {88}},\ \bibinfo {pages} {184422} (\bibinfo {year} {2013})}\BibitemShut
  {NoStop}%
\bibitem [{\citenamefont {Cranmer}(2023)}]{cranmer2023interpretable}%
  \BibitemOpen
  \bibfield  {author} {\bibinfo {author} {\bibfnamefont {M.}~\bibnamefont
  {Cranmer}},\ }\bibfield  {title} {\bibinfo {title} {Interpretable machine
  learning for science with {PySR} and {S}ymbolic{R}egression. jl},\ }\href
  {https://arxiv.org/abs/2305.01582} {\bibfield  {journal} {\bibinfo  {journal}
  {arXiv preprint arXiv:2305.01582}\ } (\bibinfo {year} {2023})}\BibitemShut
  {NoStop}%
\bibitem [{\citenamefont {Wang}\ \emph {et~al.}(2023)\citenamefont {Wang},
  \citenamefont {Hu},\ and\ \citenamefont {Sun}}]{wang2023topological}%
  \BibitemOpen
  \bibfield  {author} {\bibinfo {author} {\bibfnamefont {X.}~\bibnamefont
  {Wang}}, \bibinfo {author} {\bibfnamefont {X.-C.}\ \bibnamefont {Hu}},\ and\
  \bibinfo {author} {\bibfnamefont {Z.-Z.}\ \bibnamefont {Sun}},\ }\bibfield
  {title} {\bibinfo {title} {Topological equivalence of stripy states and
  skyrmion crystals},\ }\href {https://doi.org/10.1021/acs.nanolett.3c00718}
  {\bibfield  {journal} {\bibinfo  {journal} {Nano Lett.}\ }\textbf {\bibinfo
  {volume} {23}},\ \bibinfo {pages} {3954} (\bibinfo {year}
  {2023})}\BibitemShut {NoStop}%
\end{thebibliography}%

\end{document}